\documentclass{article}
\usepackage{authblk}
\usepackage{graphicx} 
\usepackage[left=1.5in, right=1.5in, top=1.5in, bottom=1.5in]{geometry}
\usepackage{longtable}
\usepackage{array}
\usepackage{amsmath}

\usepackage[authordate, natbib=true, backend=biber,ibidtracker=false]{biblatex-chicago}
\usepackage{booktabs}
\usepackage{longtable}
\usepackage{array}
\usepackage{makecell} 

\title{Network Inference in Public Administration: Questions, Challenges, and Models of Causality}

\author[1]{Travis A. Whetsell}
\author[2]{Michael D. Siciliano}
\affil[1]{School of Public Policy, Georgia Institute of Technology, Atlanta, Georgia, USA, travis.whetsell@gatech.edu, ORCID: 0000-0001-5395-4754 , corresponding author}
\affil[2]{University of Illinois at Chicago, Chicago, Illinois, USA, sicilian@uic.edu, ORCID: 0000-0001-8688-8685}

\date{\today}

\begin{document}
\maketitle
\begin{abstract}

Descriptive and inferential social network analysis has by now become commonplace in public administration studies of network governance and management. A sizeable literature has developed in two broad categories: antecedents of network structure and network effects and outcomes. A new topic is emerging on the analysis of network interventions that applies knowledge of network formation and effects to actively intervene in the social context of inter-organizational and inter-personal interaction. Yet, it remains an open question how scholars might deploy and determine the impact of network interventions in public administration. Inferential network analysis has primarily focused on statistical simulations of network distributions to produce probability estimates on parameters of interest in observed networks, e.g. exponential random graph models. There is less attention to design elements for causal inference in the network context, such as experimental interventions, randomization, control and comparison networks, and spillovers. We advance a number of important questions for network research, examine important inferential challenges and other issues related to inference in networks, and focus on a set of possible network inference models. We categorize models of network inference into (i) observational studies of networks, often using descriptive and stochastic methods that lack intervention, randomization, or comparison networks; (ii) simulation studies that leverage computational resources for generating inference; (iii) natural network experiments, with unintentional network-based interventions; (iv) network field experiments, with designed interventions accompanied by comparison networks; and (v) laboratory experiments that design and implement randomization to treatment and control networks. The article offers a guide to network researchers interested in questions, challenges, and models of inference for network analysis in public administration. 

\end{abstract}

\newpage 

\section{Introduction}

Network scholars in public administration have produced a variety of insights on a range of topics related to inter-organizational networks in the public sector. As \textcite{Siciliano_and_Whetsell_2023} recently argued, these insights have not yet led to a systematic understanding of what works. Practitioners and scholars alike continue to wonder, now that we understand the contours of a network, what can we do to improve their functioning? In other words, how might we use knowledge of networks to design policy and programmatic interventions to resolve public problems and enhance the general welfare? \textcite{Siciliano_and_Whetsell_2023} applied Valente's (2012) network interventions concept, developed in public health, to the context of public administration, suggesting a more systematic attack upon the what works question. \textcite{valente2012network} defines network interventions as “purposeful efforts to use social networks or social network data to generate social influence, accelerate behavior change, improve performance, and/or achieve desirable outcomes among individuals, communities, organizations, or populations” (p.49). Network intervention studies are numerous in fields outside of public administration but remain very limited in public administration. Scholars in adjacent disciplines now call for more inquiry into causal inference in networks \parencite{aral2016networked, rogowski2017causal, an2022causal}. We extend and develop these insights as a guide for public administration.

We define social networks as the complex structures that emerge from interactions between actors. This paper includes attention to both inter-organizational and inter-personal networks as the design elements and logic of inference apply to both. By inference, we mean a systematic process by which conclusions are generated from data, evidence, or facts; or as the Oxford English Dictionary defines it, “by inductive or deductive methods; reasoning from something known or assumed to something else which follows from it”. In the present context, when we intervene in networks, what are the effects of the intervention, by what processes may we know them, what common hazards should be avoided, and what practices should be employed? 

We suggest attention to five predominate categories or models of inference suitable to the analysis of network interventions: observational studies of networks using descriptive and stochastic methods that lack interventions and control or comparison networks; simulation studies that leverage computational resources to generate test environments for network interventions; natural experiments, where intervention occurs in one or more networks accompanied by analysis of a conceptually adjacent comparison network; network field experiments, where researchers actively intervene in real-world networks and compare effects to control or comparison networks; and lastly, laboratory network experiments that design and implement randomization to treatment and control networks. 

The more general purpose of this article is to stimulate interest in the epistemology of network interventions, attending to the big question of ‘how do we know what we know’ but within the context of networks. Challenges to the traditional framework of impact evaluation have arisen from the general shift toward the implementation of public policy and programs through networks. As \textcite{robins2023multilevel} note, "Intervention by necessity implicates a causal framework" (p.117). This article fills the lacuna in our awareness and understanding of the potential causal effects and challenges of developing causal insights in network contexts in public administration. 
 
\section{Methods of Network Intervention Across Disciplines}

The subject of network interventions as a topic in public administration is very new. Few articles have explicitly sought to analyze the topic, still fewer have touched on the subject of inference given network interventions. Indeed, a Web of Science search conducted in Fall of 2023 (“network intervention*” OR “network-based intervention*”) under the category of public administration returns three results: \parencite{Siciliano_and_Whetsell_2023,frank2018implementation,scott2015collaborative}. Without category restrictions, the result is 534 articles and reviews. For this reason, it is necessary to review the extant literature outside of public administration. 

\subsection{Clinical Psychology and Public Health}

The network interventions literature has origins in clinical psychology and social work research emphasizing the relevance of social networks in mental health treatment and outcomes, often referred to as ‘network therapy’ \parencite{speck1969network}. This literature begins in the late 1960s and analyzes therapeutic interventions for individuals that leverage family, friend, and broader social ties as a component of treatment. For example, \textcite{attneave1969therapy} examines therapeutic network interventions in the social context of native American tribes. Similarly, \textcite{rueveni1975network} describes a therapeutic process involving a series of network-based counseling sessions, which included extended family, friends, and neighbors. This literature largely utilizes ego-centered networks that are mapped outward from the patient. Such networks take on various types of ties, including material resource and emotional exchanges, as well as broader ‘norming’ processes and structural conditions that affect patient outcomes \parencite{hurd1981models}. Scholars here suggest network interventions fall under three model types: helping patients integrate better into existing networks, helping patients generate more robust connections, and helping patients understand their positioning within a social network \parencite{hurd1981models}. 

Later, \textcite{cohen1984network} assessed the effectiveness of network interventions in gerontology, noting the development of a standard protocol for data collection, referred to as the ‘Network Analysis Profile’. They designed an evaluation of its use in an observational study of a senior living facility. However, as \textcite{auslander1987parameters} noted, at that point in time, very few attempts had been made to measure the effects of network interventions on social support for patients. Subsequent research began to employ techniques designed to elicit causal inference. For example, \textcite{calsyn1998impact} used a randomized control trial to assess the effectiveness of network interventions for clients in the mental health and homelessness context. Clinical research showed notable improvement in the application of social network concepts, protocols for network intervention by social workers \parencite{pinto2006using}, and the application of the tools of causal inference to establish impact \parencite{soyez2006impact}. However, many studies still did not explicitly utilize ‘formal network analysis’ in terms that many contemporary network scientists would expect. Rather, these studies focused on patients and clients in a clinical setting, where networks are understood in a metaphorical, qualitative, or ego-centric manner. 

At the same time, the network intervention approach began to gain momentum in the public health literature. As a result, a stronger methodological shift occurred from clinical to inferential epistemic frameworks. For example, \textcite{el2006social} conducted a study utilizing random selection to explain the effects of network intervention on risk behaviors in the context of HIV/AIDS. Similarly, \textcite{sherman2009evaluation} and \textcite{latkin2009efficacy} analyzed the effects of peer network interventions versus life skill interventions on methamphetamine and HIV/AIDS risk behaviors using randomization and repeated measurements approach. However, while inferential methods appeared to be gaining strength, many such studies continued to be limited to the ego-centric network approach. As \textcite{latkin2015social} suggested, the literature focused primarily on ‘personal’ networks with fewer articles branching into ‘sociometric’ networks. 

In 2012 Thomas Valente published “Network Interventions” in \textit{Science}, which provided a descriptive typology of network intervention methods directly leveraging tools from network science. As he suggested, network interventions tend to fall into four categories: individual, segmentation, induction, and alteration. The clinical approach to helping patients exemplifies the individual approach, while the public health approach to targeting segmented vulnerable groups and inducing network processes tends to exemplify both the segmentation and the induction approaches. Less research has been conducted using the alteration approach. The alteration intervention requires explicit information about network structure in order to make informed choices about which nodes or ties to target for alteration, i.e. addition and deletion. Further, Valente's (2012) conceptualization reflected the increasing necessity to go beyond metaphorical use and engage in the collection and formal analysis of social network data. 

The approach suggested by \textcite{valente2012network} further coincided with newer developments in inferential social network analysis, which included the development of exponential random graph models (ERGMs) and stochastic actor-oriented models (SAOMs) \parencite{lusher2013exponential, snijders2010introduction}. For example, \textcite{steglich2012actor} used SAOMs to estimate the social network determinants of smoking in school children with implications for smoking cessation interventions. Similarly, \textcite{haas2014little} conducted a study on the asymmetric effects of peer influence on starting and quitting smoking in adolescents, using the SAOM approach. Further, \textcite{henry2016analyzing}estimated the effects of physical activity on friendship formation among school children suggesting important context factors for interventions targeting exercise using ERGMs. These inferential network analysis approaches combine observational data with computational elements of simulation, which are used to generate inferences about interventions \parencite{cranmer2020inferential}. One study by \textcite{el2013network} employed a simulation approach to study network interventions among well-connected nodes in the context of obesity.

Further building on the causal evaluation of network interventions in the public health setting, \textcite{hunter2019social} conducted a retrospective meta-analysis of 37 studies. They found 27 RCTs and several other studies with varying designs, including partial RCTs, cluster RCTs, randomized placebo-controlled designs, randomized crossovers, controlled before and after designs, and quasi-experimental designs. They categorize every study by Valente's (2012) taxonomy, finding that the majority use individual and induction strategies. While the outcomes of these studies are not of interest here, \textcite{hunter2019social} note that treatment contamination across groups was a biasing factor. They raise an important point with respect to inference in network studies: “the correct way to obtain unbiased effect estimates is not obvious and certainly nontrivial when individual observations fail the independence assumptions that are required for conventional analysis”(p.16), a point which will be discussed at length in a later section. 

\subsection{Organization Theory and Organizational Behavior}

As \textcite{Siciliano_and_Whetsell_2023} observed the vast majority of network interventions research focuses on inter-personal networks and has ignored institutional determinants of network structure and outcomes. The intervention literature also had little to say about the world of inter-organizational. Thus, there was no development in research questions such as how do policies and programs affect the formation or dissolution of ties in governance networks, what are the consequences for outcomes of interest in governance networks when collaboration structures are altered, and how do we assess impact with respect to policies and programs implemented in an inter-organizational network context? The key distinction here between the clinical psychology, public health, and even to some extent the economics literatures and the public administration literature on the topic of network interventions is the organizational valence of the latter, where the relevant network is either within organizations or between them. 

As a consequence, techniques of network inference (observational, simulation, field experiments, and experimental designs) have been almost entirely developed with respect to individuals. This is unfortunate and somewhat remarkable considering the simultaneous flourishing of research in public administration under the banner of network governance \parencite{siciliano2021mechanisms, medina2022network}. While studies in this vein uncovered a variety of network elements, leveraging myriad concepts and methods, much, if not all, of this work remains fundamentally descriptive. We have avoided questions of causal inference regarding the effects of policies and programs on governance networks. Hence, we remain unprepared to fully address the practical question of what works. 

Early work lamented the lack of a theory of network formation \parencite{hay2000tangled} and sought to build up a theory of social exchange, emphasizing shared agendas and pooled resources. \textcite{toke2000policy} sought to build up a formation theory on individual cognitive factors through qualitative analysis of discourse. \textcite{graddy2006influences} added to the literature by identifying inputs to network size and scope, such as organizational size and resource dependence, using traditional statistical methods. Much of the early literature here treated the network either in a metaphorical way, ‘networking’, or in an ego-centric approach counting the number of partners in a focal organization’s portfolio \parencite{meier2003public}. Such studies did not gather any actual social network data or employ any kind of formal social network analysis. At this point, as a discipline, we were at least 30 years behind developments in parallel disciplines. For example, \parencite{krackhardt1988predicting} and \textcite{krackhardt1988informal} had been conducting formal network studies within organizations since the 1980s. As shown earlier, clinical psychology and social work have been implementing network interventions since the 1970s.  

Provan and Milward (1995, 1998) lamented the treatment of networks as metaphor, and also as managerial networking early on, suggesting attention to formal network structure. Much of this work provided a good foundation for the development of network governance. Yet arguably, it was overshadowed by the metaphors of network management until an attempt was made to advance a theory of network effectiveness vis \textcite{provan2008modes}. This landmark work connected distinct policy domains to distinct ‘modes’ of network governance, representing a contingency theory of network effectiveness. 

Around the same time, more rigorous applications of network analysis began to be published in public administration. \textcite{henry2011ideology} conducted an analysis of the determinants of network structure, gathering social network data and leveraging a network correlation method similar to the quadratic assignment procedure (QAP). \textcite{berardo2010self} applied exponential random graph models (ERGMs) to estimate the probability of ties forming in water management networks, which were also suited to analyzing dyad dependent processes in networks. \textcite{lee2012interorganizational} used ERGMs to analyze formation in regional economic development networks. \textcite{scott2015collaborative} used ERGMs for environmental restoration networks. \textcite{siciliano2015advice} applied ERGMs to the intra-organizational context to analyze determinants of network formation among schoolteachers. As these studies rightly noted, using the traditional methods of statistics on network data risked biasing estimators due to dependence in relational data \parencite{robins2012statistical}. What they further revealed was the effect of a complex set of endogenous drivers of network formation that had not been fully addressed in public administration, including homophily, transitivity, and preferential attachment \parencite{siciliano2021mechanisms}.

Nevertheless, there remains limited work that seeks to establish causal inferences about public action upon networks. For example, \textcite{scott2015collaborative} study whether government sponsorship of collaboration enhances network formation using ERGMs. \textcite{whetsell2020government} framed a study of the implementation of a network administrative organization (NAO) as a catalyst of network formation, using longitudinal network data with ERGM and SAOM models to estimate NAO effects. As far as we aware, no published studies exist in our field that have explicitly designed and implemented a public policy or program to change the shape or structure of an inter-organizational network. In public health, randomized control trials of network interventions are now so numerous that meta-analyses may be conducted on them as a group. There are no doubt good reasons for the differences, such as cost and authority. But perhaps also because public administration has so little experience in estimating their potential effects. In other words, perhaps we have just aimed too low. How might such studies be structured and what kinds of hazards might they encounter?   

\section{Causal Questions for Networks in Public Administration}

Before exploring in more detail the tools and approaches one may use to estimate causal effects with network data, let’s first (i) clarify the types of causal questions public administration and policy scholars may study and (ii) highlight the challenges associated with answering these questions.  To begin, it is helpful to note that network-based questions are defined by the level of analysis \parencite{siciliano2022networks, borgatti2022analyzing, Krackhardt2010encyclopedia}. Most network questions reside at one of three levels: the network level, the node level, and the dyad level. These levels are relevant to any network, regardless of the actor type (people, organizations, countries) or relationship type (friendship, advice, formal contracts). Therefore, it is important to distinguish the unit of analysis from the level of analysis.  

The units of analysis are the types of nodes that comprise the members of the network. So if our nodes are school teachers, then school teachers are the unit of analysis. If our nodes are local governments, then local governments are the unit of analysis. The level of analysis, on the other hand, concerns the structural feature of the network that one is interested in modeling or explaining. To make this discussion more concrete, let’s assume we have a network of 10 cities in a region, and the ties among them are shared service agreements related to sustainability. At the network level of analysis, we are interested in the composition and structure of the network as a whole. Perhaps, one wants to explore the centralization or density of the network. For these measures, we have only a single observation, as network-level measures describe the entire network. 

At the node level of analysis, we are interested in the structural positions and performance of the nodes rather than the network as a whole. For example, we may want to examine how a city's fiscal stress influences its position in the network, perhaps as measured by betweenness centrality. In this instance, we would have 10 observations, one for each of the cities in the network.

At the dyadic level of analysis, our interest lies in the presence and absence of the ties among the nodes. One may be interested in explaining why service contracts are formed between certain cities but not others. Because each of the ten cities could possibly have a service contract with nine other cities, the number of observations at the dyadic level is 90. As evident in this brief discussion, the level of analysis for a given research question dictates the number of observations one has available to answer that question. For our example network with 10 cities, we had one observation at the network level, 10 at the nodal level, and 90 at the dyadic level (assuming the relations are directional and 45 if the relations are symmetric). The level also determines the amount of dependency and autocorrelation in the data. An issue we discuss in more detail later. 

In addition to the level of analysis, network research is often divided into two major branches: antecedents of networks and consequences of networks \parencite{borgatti2003network, siciliano2022networks}. The antecedents of networks focus on the various factors and mechanisms that facilitate or constrain the formation of ties among the actors. Common theories and mechanisms of tie formation include homophily (i.e., actors who share similar traits are more likely to form relationships), transitivity (i.e., the tendency for actors who share a common third party to also form a relationship), and preferential attachment (i.e., the tendency to form ties with actors who are popular) \parencite{siciliano2021mechanisms}. 

Research focused on the consequences of networks treats the network and its composition and structural features as the independent variable. Thus, the emphasis is on understanding the implications of certain network configurations for the actors and system as a whole. Common mechanisms and theories used to explain the outcomes associated with networks include social capital (i.e., the benefits to actors based on bridging and bonding), centrality (i.e., how the structural position of an actor affects its success), and density (i.e., how the overall connectivity of the network affects its performance) \parencite{medina2022network}.  Thus, network research is interested in two forms of causation \parencite{vanderweele2013social}. The first form is on the various factors, contexts, and actor attributes and behaviors that shape the formation of ties. The second form of causation is focused on the consequences of network structure and how actor positions and relationships influence the attributes and behaviors of those in the network. Overall, researchers in public administration are interested in explaining both the antecedents and consequences of networks at different levels of analysis. Table 1 summarizes these distinctions and provides a representative causal question. For this table, we will stick with the example above of a regional network comprised of cities working together on sustainability issues. One could imagine gathering data on multiple regions as needed by the research question to gain variation on network-level variables.

\begin{longtable}{>{\raggedright\arraybackslash}p{0.25\linewidth} >{\raggedright\arraybackslash}p{0.35\linewidth} >{\raggedright\arraybackslash}p{0.35\linewidth}}
\caption{Causal Questions for Network Levels} \label{tab:my_label} \\
\toprule
\textbf{Level of Analysis} & \textbf{Network Antecedents} & \textbf{Network Consequences} \\
\midrule
\endfirsthead
\toprule
\textbf{Level of Analysis} & \textbf{Network Antecedents} & \textbf{Network Consequences} \\
\midrule
\endhead
\bottomrule
\endfoot

\textbf{Network Level} & 
\textbf{Question:} Does the creation of a regional institution dedicated to sustainability (such as Chicago’s Greenest Region Compact) foster greater collaboration among local governments in that region?\par\vspace{2mm}
\textbf{Treatment:} Presence or absence of a regional institution.\par\vspace{2mm}
\textbf{Outcome:} Level of regional collaboration. This could be measured by the density of the collaboration network. & \textbf{Question:} Do regions with greater levels of collaboration produce better sustainability outcomes (e.g., reduced greenhouse gas emissions)?\par\vspace{2mm}
\textbf{Treatment:} Increase in level of regional collaboration.\par\vspace{2mm}
\textbf{Outcome:} Improved sustainability outcomes for the region. \\
\addlinespace
\textbf{Node Level} & 
\textbf{Question:} Are local governments with a dedicated sustainability officer more central in the regional collaboration network?\par\vspace{2mm}
\textbf{Treatment:} Presence or absence of a dedicated sustainability officer.\par\vspace{2mm}
\textbf{Outcome:} Centrality of the local government in the regional collaboration network. & \textbf{Question:} Are local governments with more bridging relations more likely to meet their sustainability goals?\par\vspace{2mm}
\textbf{Treatment:} Increase in bridging relations.\par\vspace{2mm}
\textbf{Outcome:} Improved sustainability outcomes for the local government. \\
\addlinespace
\textbf{Dyadic Level} & 
\textbf{Question:} Are two local governments that participate in a regional forum (e.g., council of government) more likely to collaborate with each other?\par\vspace{2mm}
\textbf{Treatment:} Mutual participation in a regional forum.\par\vspace{2mm}
\textbf{Outcome:} Formation of a collaborative tie. & 
\textbf{Question:} Are local governments that collaborate on sustainability issues more likely to set similar sustainability goals (e.g., transitioning government vehicle fleets to electric)?\par\vspace{2mm}
\textbf{Treatment:} Presence or absence of a collaborative tie.\par\vspace{2mm}
\textbf{Outcome:} Similarity in sustainability goals. \\
\end{longtable}

Each of the questions in Table 1 deals with important inquiries about how government institutions should be designed and incentivized to create strong collaborative networks and to understand the implications of those networks on outcomes of interest. While this example centered on local governments engaged in sustainability activities, these questions could easily be framed to consider other relationship types and policy domains as well as intra-organizational networks (e.g., police officers in a police department or teachers within a school). As noted above and by \textcite{Siciliano_and_Whetsell_2023}, causal questions about what works regarding network formation and network performance have not been addressed in the fields of public administration and policy. 

As may be evident to readers familiar with social networks, some of the proposed treatments are difficult to apply in a random or exogenous manner. This is a key reason why much of the research addressing questions similar to the ones posed above have been based primarily on descriptive and associational analyses. Take the example of network-level consequences concerning how network density may improve sustainability outcomes. A critical challenge is determining how to increase density in an exogenous fashion. What policy levers or changes in rules or norms might influence tie-formation decisions to increase connectivity? Such manipulation of network structures might be difficult to create outside of the laboratory (for examples of lab-based studies, see: \textcite{Bavelas, leavitt1951some, centola2010spread}). Or consider the question relating the bridging ties of a local government to its performance. Exogenously increasing bridging ties for some actors, but not others may be difficult, though possible (see, for example, \textcite{carnabuci2023people}). \textcite{Siciliano_and_Whetsell_2023} provide a review and expansion of network intervention strategies applicable to public administration and policy. 

A major goal of this article is to encourage scholars in public administration and policy to pursue and adopt stronger research designs.  Such designs and strategies provide stronger causal claims and evidence of what works. Before considering different methods and strategies for network inference, we first need to consider the major challenges to making causal claims with network data.

\section{Challenges for Causal Inference}

Answering causal questions about networks presents a number of methodological issues. In this section, we will highlight some of these challenges. Several of the challenges are unique or more acute in network studies, while others are present in most empirical research.  

\subsection{Selection Bias}

One of the key challenges, especially in public administration, arises from the fact that the majority of network-based studies rely on observational data \parencite{siciliano2022networks}. Consequently, researchers lack experimental control and exogenous variation on the variable of interest. This leads to concerns about selection bias. In the potential outcomes framework, considering a binary treatment, selection bias is defined as:

\begin{equation}
    \text{Selection Bias} = E[Y^0 \mid D=1] - E[Y^0 \mid D=0]
\end{equation}

Selection bias is the difference in expected outcomes between the treated, D = 1, and the untreated, D = 0, groups, if no treatment was given to either, Y0 \parencite{cunningham2021causal}. In other words, selection bias is the difference in the outcomes we would expect between the two groups in the absence of treatment.  A classic example is the question of the difference in standardized test scores between children who attend private school and those who attend public school. Because families with certain characteristics self-select into having their children attend private school, we can’t simply compare the outcomes between public and private school children. Those children are different in observable and unobservable ways. The amount of selection bias present would be the difference in standardized test scores between private and public school children had both groups of children went to public school. 

We typically think of selection concerns with regard to treatment status, such as attending a private school or participation in a job training program. These same concerns hold in network settings. For example, if one was interested in exploring variation in network size and structure between private school and public school students, we can’t consider school status as exogenous. In addition, with network data, we are also concerned with selection into specific ties and network positions. For example, a large body of research has found that individuals who play bridging roles in networks tend to have higher levels of performance, innovation adoption, and pay \parencite{burt2000network, burt2009structural}. But because individuals with certain attributes and personalities may be inclined toward bridging positions, concerns of selection bias are warranted. Thus, with regard to networks, network effect models may be biased to the extent that individuals select into their network positions. \textcite{li2013social} cites a revealing comment by \parencite{fleming2007brokerage}. As they state: “network analysis has mainly ignored the problem of separating the causal effects of position from the intentions of the individual who may have consciously and strategically created the position”(p.177).

\subsection{Omitted Variable Bias}

Omitted variable bias is a common concern for most empirical analyses. As widely discussed in econometrics textbooks \parencite{wooldridge2008introductory}, the coefficient on the variable of interest in a regression model will suffer omitted variable bias if two conditions are met. First, the variable of interest has to be correlated with the excluded variable, i.e., the one omitted from the model. Second, the variable that was excluded must be correlated with the outcome of interest. Though less often discussed, network-based methods of inference, such as ERGMs and SAOMs, also suffer from omitted variable bias. Therefore, researchers need to carefully consider the data-generating process and include all relevant variables. For example, suppose one is interested in the effect of transitivity on tie formation in a network, but leaves out an important homophily term. In that case, the coefficient on transitivity will be biased as clustering in networks is also driven by homophily. 

More generally, in network models, there are two common variable types that raise concerns with omitted variable bias: shared contexts and shared relations. Shared contexts are the common environments and settings in which people interact. Shared contexts include work and school environments, social institutions (such as churches, soccer leagues, etc.), and geography.  These shared contexts can influence both the likelihood of two actors forming a relationship as well as their attitudes and behaviors. If relevant shared contexts are omitted, the coefficients of interest may be biased.

Omitting shared relations is the second variable of concern. All network studies need to bound the network in some fashion. The bounding process can omit key actors from consideration. For example, a study of teachers may bound the network based on the school in which one works. But if teachers form meaningful ties with peers in other schools, then it is likely that these ties shape not only their behavior in the school but also the ties they form. Regarding the latter, models of network formation based on a bounded network will be misleading if the omitted actors have ties with members of the bounded network. For a simple example, consider Figure 1. The large circle bounds the network, and a single node, actor ‘E’ has been left out. ‘E’ has ties with the actors in the network, but those ties are not witnessed as ‘E’ was excluding from the study. Based on the bounded network, one might conclude that there is no clustering. But that conclusion would change if actor ‘E’ was considered. 

\begin{figure}[ht]
    \centering
    \includegraphics[width=0.25\textwidth]{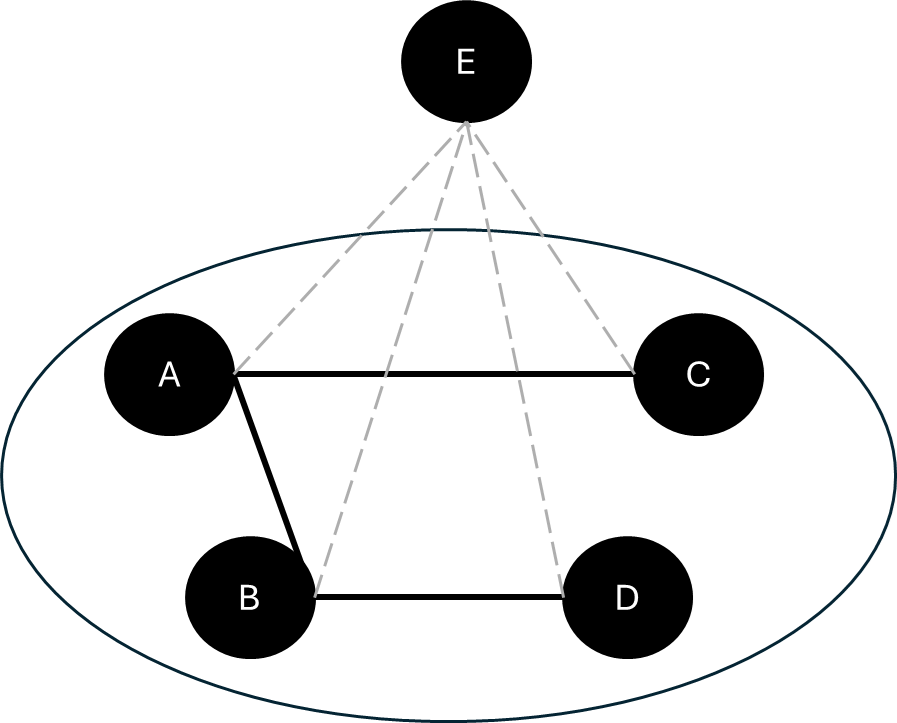} % Adjust the file name and path as needed
    \caption{Omission by Network Boundary}
    \label{fig:my_label}
\end{figure}

\subsection{Non-independence of Observations}

One of the major underlying assumptions of standard statistical analysis is that the observations are independent and identically distributed (IID). The IID assumption states that (i) the outcome or behavior of one observation does not influence the outcomes or behavior for another and that (ii) the observations come from the same probability distribution. Random sampling is a key method to obtaining data that is IID. However, for most network studies, random sampling is not possible. It is not possible because we are interested in collecting data on the entire population of actors in the network of interest.

Concerns about non-independence primarily affect nodal and dyadic level analyses. At the nodal level, the primary concern is network autocorrelation. Network autocorrelation arises when attribute or outcome values for nodes are similar to those of their alters. This similarity can occur through a variety of forces, and (especially with cross-sectional data), it can be very difficult to determine which is driving the autocorrelation \parencite{shalizi2011homophily}.  Three forces to consider are contagion, homophily, and shared contexts/environments. Contagion, or social influence, is the process by which ideas, attitudes, and behaviors spread through social interaction \parencite{friedkin2011social}. Actors who are connected to one another may begin to think and act in a similar manner due to their interactions.  The issue of contagion will also be discussed under SUTVA violations. Homophily refers to the tendency for similar actors to seek relationships with one another \parencite{mcpherson2001birds}. Tendencies toward homophily have been well-documented in a variety of settings from youth friendships \parencite{goodreau2009birds} to organizational networks where ties are driven by similar missions and funding sources \parencite{atouba2015international}.

Similarly, at the dyadic level, non-independence and autocorrelation is a very significant concern. Autocorrelation at the dyadic level arises through several processes. Consider a directed network represented as an adjacency matrix, where the ‘i,j’ cell indicates the presence or absence of a relationship between actor i and actor j. In the matrix, all of the cells in row i have actor i as the sender. And all of the cells in column j have j as the receiver. So any attribute about i or j that makes them more outgoing or attractive leads to correlation among the dyads containing that actor. More problematic, however, are higher-order structural dependencies. The dyads in the network are not only dependent upon the actors who comprise each dyad (e.g., actors i and j) but potentially all other dyads in the network, which is a primary reason that standard statistical models are inappropriate in the network context. The old adage ‘a friend of a friend is a friend’ is a common example. Two actors in a network are much more likely to become friends, if they have a friend in common. Thus, we can’t simply look at actors i and j to assess the likelihood that they are connected, we need to know the status of their relationship with all other actors in the network. A number of network models have emerged that attempt to account for or directly model these higher-order dependencies. These network models include the quadratic assignment procedure (QAP) \parencite{krackhardt1988predicting}, ERGMs \parencite{lusher2013exponential, robins2007recent}, and stochastic actor oriented models (SAOMs) \parencite{snijders2017stochastic, snijders2010introduction}. 

At the network level, the assumption of independence becomes much more tenable. For example, if we gather network data on local government collaborations for sustainability in different metropolitan regions around the country, we could argue that each regional network operates in relative independence from one another. While independence can be gained, the data collection demands for statistical analysis when the network is the level of analysis are much higher. 

\subsection{Spillovers and SUTVA Violations}

Observations of interest in a network are not independent, and therefore random assignment to treatment groups within a network violates standard SUTVA assumptions. SUTVA stands for the Stable Unit Treatment Values Assumption. SUTVA requires that the outcome observed for one actor does not depend on the treatment assigned to other actors. In other words, there are no spillover effects or network interference \parencite{leung2020treatment, bhattacharya2020causal}. Consider a medical trial where patients are randomly assigned to take an experimental drug or a placebo. Because those who participate in the trial are assumed to be independent of one another and because the treatment assignment of one participant has little influence on the potential health outcome of others, SUTVA is usually thought to hold in these settings \parencite{hong2013heterogeneous}. 

However, SUTVA is violated in networks because the actors are interdependent and influenced by their relationships with others. Consider a situation where several local governments out of dozens in a region each agreed to hire a sustainability director. Assume a study was interested in testing whether the presence of a director improves the sustainability initiatives of the “treated” local governments. Because the local governments are all part of the same region and interact to coordinate and address regional challenges and sustainability issues, SUTVA is violated. Indeed, if we consider network formation and network effects, both types of studies will be affected. On the network formation side, the treatment status of actors in the network may influence the presence and absence of the social ties of others. For example, if those cities with a sustainability director become more active in the network, it is not only their ties and position that is changing, but that of the other governments in the region as well. On the network effects side, social influence may operate whereby the behaviors of one actor may influence those adopted by others. Perhaps those who hired a sustainability director tended to set more ambitious sustainability goals and such decisions may influence the goals set by other cities in the region. 

To overcome these issues, cluster-randomized trials have been proposed and used in many settings where participants interact \parencite{hong2013heterogeneous}. In cluster-randomized trials, every member of the cluster is assigned to the same treatment.  These techniques have been widely deployed in education settings (where entire schools are randomly treated) \parencite{hong2013heterogeneous} and development economics (where entire communities are randomly treated) \parencite{banerjee2010improving, crepon2013labor}. Such approaches translate nicely to network settings, where rather than randomizing treatment at the node level, treatment is assigned at the network level.

Research on peer effects, which have been one of the most studied dynamics associated with networks \parencite{bramoulle2020peer}, is in fact a study of interference and SUTVA violation. A motivating question in the network literature has been to try an understand how the treatment given to actor i has a causal effect on the outcomes for actor j. Thus, whereas causal inference in other fields requires SUTVA to hold, network processes and dynamics are of interest because of SUTVA violations. 

\subsection{Other Considerations}

There are several other issues to consider as well. These include simultaneity, measurement error, and mechanism identification. Simultaneity presents a challenge for most empirical research. In a typical regression model, concerns with simultaneity arise when the predictor variables and the outcome variable mutually influence one another. A classic example from economics is supply and demand. In network settings, concerns with simultaneity arise in at least two scenarios. First, studies that seek to associate an actor’s position with outcomes need to clarify that it is indeed the structural position that leads to a given outcome, rather than the presence of the outcome leading to a particular position \parencite{li2013social, an2022causal}.  Second, in models exploring social influence, individuals in the network are mutually influencing one another making it difficult to discern the direction. 

The role of measurement error in biasing parameter estimates is well established in econometrics \parencite{wooldridge2008introductory}. Measurement error in the explanatory variables causes attenuation bias and can lead to incorrect inferences. In relation to network data, measurement error is a primary concern with regard to how accurately one is able to capture the social structure of interest. Studies over several decades have shown that people often have poor memories of their social interactions and vary in their ability to accurately recollect them  \parencite{bernard1982informant, killworth1976informant, ertan2019perception, casciaro1999positive}.  

Finally, there is the need for and challenge of identifying the specific mechanism that may be at work. Even in the area of peer effects, which have been the most widely researched network dynamic from a causal inference perspective, the particular mechanisms that drive the observed peer effects remain uncertain. As \textcite{bramoulle2020peer} note: “A well identified estimation of peer effects only constitutes a first step in the analysis of social interactions. Peer effects can have different causes, including complementarities, conformism, social status, social learning, and informal risk sharing.”(p.611)

\section{Models of Network Inference}

Literature on causal inference tends to distinguish between observational studies and experimental studies, where observational studies lack randomization of subjects to control and treatment groups and where an intervention is designed and implemented to elicit some outcome that has effects in one group but not the other. The purpose of such control is to establish inferences about the intervention relative to a counterfactual reality where the intervention did not take place \parencite{pearl2009causality}. However, there is variety within the observational to experimental continuum, often termed 'quasi-experimental'\parencite{cook2002experimental}. The relevant categories presented here include 1) observational network studies, 2) network simulation studies, 3) natural network experiments, 4)  network field experiments, and 5) laboratory network experiments. These categories are marked by variation on 1) the type of network intervention, where some studies explicitly design network-based interventions, and 2) the type of experimental control, where studies implement randomization or attempt to establish valid comparison networks. 

\subsubsection{Preliminary Notes on Models of Network Inference}

To generate network data the researcher must identify both nodes and the relational tie. For example, a survey may ask the respondent to generate a list of names with whom they interact. This is often difficult since individuals tend to not want to “name names” or may have difficulty with recall. However, it is critical to gather such information to conduct formal network analysis. This is often referred to as the name-generator approach \parencite{burt1984network}. Another similar approach is the roster method. If the researcher then iterates across actors within a network boundary they can generate a tie list sufficient to construct a network \parencite{laumann1989boundary}. Network data identifying both nodes and ties can also be gathered from documentary sources that do not rely on subjective recall. These include agreements, alliances, or partnerships between organizations; board membership lists gathered from tax forms; email communication records; co-participation in the meeting minutes of collaborative forums; or joint participation on planning documents. 

Assuming the researcher can gather basic information on nodes and ties and construct a network, such networks may be composed of varying types of network ties. For example, common types of ties between actors are communication \parencite{lee2018does} information \parencite{whetsell2021formal} or advice seeking \parencite{siciliano2015advice}. Another example of a tie commonly found in public administration is resource-exchange in inter-organizational networks \parencite{whetsell2020government}. For example, public, private, and non-profit organizations often engage in a complex arrangement of partnership ties, where information, resources, human capital, etc. are exchanged. Aggregating nodes and ties often reveals a complex structure of dyads, triads, chains, cliques, and various topological features. 

\subsection{Observational Network Studies}

The first and most common type of network study design in public administration is the observational study. These tend to be single network cross-sections without network interventions and without control or comparison networks. Sometimes observational network studies collect temporal network data but lack experimental intervention or control networks. Such studies tend to focus on correlations between variables, and analysis is accomplished through a variety of methods \parencite{siciliano2022networks}. 

The observational network study is one in which the researcher can identify network ties where nodes have unique identifiers and ties refer to a specific relationship between them and then aggregate these together to establish network(s) which may be analyzed at the whole network level, the dyad level, or the individual node level. However, such studies tend to lack experimental intervention and control and suffer many of the same threats to validity that other observational studies are subject to, in addition to the specific threats encountered in the network setting identified above. As such, one may be limited in the causal inferences that may be derived. Despite the limitations of observational network studies, they continue to be of great importance to the development of network theory and methodology in the discipline. 

Many network researchers in public administration are interested in describing the landscape of interactions between actors, and typically identify whole network statistics, such as centralization, while also generating node position statistics such as node centrality \parencite{nowell2019networks, nowell2023population}. Others may be interested in estimating the effects of statistics derived from networks, such as node centrality, on other outcomes, such as performance, within the context of the standard linear model. Such studies are often referred to as ‘network effects’ studies \parencite{medina2022network}. Generally, network effects studies tend to establish correlations rather than estimating effects of network interventions in an experimental manner. 

Researchers may also be interested in identifying the antecedents of ties, i.e. what processes and variables lead individuals and organizations to form relationships \parencite{siciliano2021mechanisms}. It has become standard to employ exponential random graph models (ERGM) to estimate tie formation effects. Such models allow researchers to identify so-called ‘endogenous’ processes that operate in a wide array of social, biological, and physical networks. These include, for example, reciprocity \parencite{gouldner1960norm}, transitivity \parencite{girvan2002community}, and preferential attachment \parencite{barabasi1999emergence}, which serve as general processes responsible for generating ties across the network.Extensions of these models exist for temporal network data. For example, \textcite{ingold2016structural} compare cross-sectional ERGMs to (T)emporal ERGMs and (S)eperable TERGMs to analyze perceived influence in policy networks; \textcite{yoon2023antecedents} uses TERGMs to analyze board inter-locks in the non-profit sector; and \textcite{liu2024member} use TERGMs to analyze the influence of managers in interlocal agreements.

 There are also dyadic processes such as assortative mixing, i.e. homophily and heterophily, which facilitate ties between nodes based on shared (homophily) or different (heterophily) attributes \parencite{mcpherson2001birds}. Researchers often collect a range of attribute data on the nodes of interest, such as age, size, motivation, sector, etc. In addition to considering assortative mixing, these attribute variables can also be utilized as 'exogenous' predictors of tie formation. For example, in an ERGM setting, one could look at sender and receiver effects to address questions of whether public, private, or non-profits sector organizations are more likely to form ties in a given network \parencite{lusher2013exponential}. While ERGMs are considered a class of inferential network models, they may be viewed through the causal lens in a similar manner to standard regression models: the research design used to generate the data determines whether or not causal statements are warranted not the method of analysis. 

\subsection{Network Simulation Studies}

Network simulation studies often specify the rules or processes of interaction between nodes and permit their manipulation in order to observe differences in network outcomes. These studies often dictate that certain node types are more likely to form ties while others are less likely. Following the framework proposed by \textcite{valente2012network}, they may alter networks, adding and removing nodes, or adding and removing ties, in order to observe changes in network processes and structures. They may incorporate shocks dynamically into a system, for example, modeling a disaster response by a group of cross-sector network actors \parencite{li2019modeling}. Such studies may also simulate a program or policy intervention designed to stimulate or suppress network activity. 

Simulation studies may have a few advantages over observational studies. For example, such studies can simulate interventions, control/comparison networks, and their dynamics over time. However, simulation studies are only as good as the assumptions and rules that structure them \parencite{desai2012simulation}. An advantage is that network simulation may start with assumptions derived from observed network cross-sections but then incorporate such assumptions into the design rules of the simulation. 

In addition, methodological developments in inferential network analysis, e.g. stochastic actor oriented models (SAOM), in addition to providing options for longitudinal network models and co-evolution models also permit robust simulation capabilities and may be empirically calibrated with reference to existing data sets \parencite{steglich2022stochastic}. The ERGM framework also provides simulation capabilities in the network setting. For example, networks can be generated based on observed network data, and initial parameters on processes such as reciprocity or transitivity may be adjusted to simulate potential impacts. Similarly, nodes may be added or removed to simulate effects on network processes and structure.  

There currently exist only a handful of simulation-based studies in public administration. One example is \textcite{scott2019convening}, which combines network analysis with agent-based models to simulate differential effects on collaborative processes. Similarly, \textcite{bitterman2020modeling} employ simulation techniques to test the effects of collaborative governance design on performance outcomes. \textcite{lazer2007network} conduct a simulation combining agent-based models and network analysis to generate theoretical propositions about the effects of communication patterns in problem solving networks on organizational performance. \textcite{koliba2019using} discuss the use of agent-based models and simulation techniques in public administration more broadly. 

More generally, simulation studies are often very useful for decision-making but are also useful in generating inferences in the face of complexity. This is becoming truer as computational power continues to develop along the trajectory roughly approximated by Moore’s Law. Studies in the future may be able to generate simulation rules derived from vast databases of network data. 

\subsection{Natural Network Experiments}

Conditions frequently arise in public administration where one set of actors experiences the benefits of a particular policy or program while another does not. Often policy interventions are crafted without a conscious understanding of networks, but such interventions nevertheless have network-based effects. For example, industrial policy targeting a particular sector may seek to enhance or inhibit the development of cooperative relationships between public, private, and non-profit organizations \parencite{mowery1998changing} without reference to broader network processes or structures.

Government intervention in the technology sector is particularly relevant \parencite{bozeman1994evaluating, bozeman2000technology, salter2001economic}. Nations compete globally to develop and maintain ‘strategic’ domestic markets that do not rely too heavily on foreign partners or investment. These often take an economics approach, but public administration is implicated when an inter-organizational approach is taken. For example, in the former approach, policymakers may seek direct subsidies or sanctions that target individual firms, but in the latter, a network approach may target cooperation between domestic organizations as a source of national competitive advantage \parencite{browning1995building}. Here, key concepts from network governance, such as the 'network administrative organization' become relevant \parencite{provan2008modes}. 

Such situations may be reconceptualized for inference in terms of natural experiments. In these cases, it is important for the researcher to gather relational network data across not only the entire target network but also conceptually adjacent networks. This has been achieved in the economics and business management literature. For example, the literature on strategic alliances has long recognized the importance of networks, and numerous articles compare networks across markets \textcite{rosenkopf2007comparing, schilling2009understanding}. However, rarely do such studies implement network interventions, and randomization remains difficult outside of laboratory settings. 

To build off one example in public policy, \textcite{scott2015does} applies matching to establish comparisons between treated and untreated areas, where the intervention is the presence of a collaborative forum and the outcome is environmental quality. Such a design could be applied with the added context of network data to create a natural network quasi-experiment design. A hypothetical example might be to compare the effects of programs that support local resilience planning networks in one geographic location versus another location that receives no such support. Another example might be to compare the effects of legislation that mandates participation of public organizations in collaborative forums in one jurisdiction versus another with no such mandate. 

\subsection{Network Field Experiments}

The key distinction between natural and field experiments is that the research actively intervenes in the 'treatment' network, whereas in the natural experiment, the 'treatment' is implemented by the policymaker without considering network effects. Additionally, keeping the inference problem of spillovers in mind, network researchers need to put extra effort into crafting control and comparison networks that are not affected by treatment networks.

Network field experiments have been conducted more widely outside of public administration, having rarely been conducted on public-oriented networks with an eye toward public problem resolution. For example, \textcite{hasan2015peers} estimate the effects of a college program randomizing peer interaction on social network growth and measure the effects by conducting successive waves of network surveys before and after the peer intervention. Similarly, \textcite{carrell2013natural} evaluate the effect of randomizing network composition on academic performance, through assignment to control and treatment networks. Another interesting example, is \textcite{lungeanu2023tale} who conducted a field experiment on astronauts training for deep space travel to to determine the effects of isolation on team structure. While the isolated crew executed its mission in a controlled setting, the researchers constructed comparison teams of university students to perform similar tasks in a non-isolated setting and took repeated measurements of network structure for both over time.  

One hypothetical approach at in the public administration setting at the individual level might be to implement collaborative interventions within organizations in separate departments or geographically disbursed units while monitoring inter-personal network activity in both treatment and comparison networks before and after the intervention. 

\subsection{Network Laboratory Experiments}

Network-based laboratory experiments establish experimental network control and conduct network interventions, typically in a physical and/or computer environment. Such experiments have the greatest internal validity and control for threats common in network settings more effectively. Though, concerns with external validity are often more pronounced.

While network-based experiments are quite rare, they are becoming more common. For example, \textcite{shore2015facts} conducted a laboratory experiment where participants were assigned to networks of varying structure and played a 'whodunit' type game. The experiment was designed to test the effects of network structure on exploration behavior. Another example, is  \textcite{mcbride2013enemy} who conducted a laboratory experiment where subjects were shown partial network diagrams of fictional dark networks of illegal activity and then made to choose which nodes to arrest in order to determine optimal network disruption decisions. Similarly, \textcite{mcbride2013efficacy} conducted a laboratory experiment where subjects are shown network diagrams or table format data in order to make network disruption decisions. Shifting topics, \textcite{takacs2018referrals} conducted a laboratory experiment to test whether referrals and information flows increase discrimination in the hiring process. 

To our knowledge there have been no published studies of laboratory-based network experiments in public administration journals. Such studies would bring participants into the laboratory and randomly assign individuals to a treatment versus control network to assess the effect of some intervention strategy and isolate any spillover effects between the networks. 

A hypothetical example might be something like the following. Individuals are recruited from work in various levels of government, non-profits, and the private sector, and depending on the research question, they could be assigned randomly to a ‘treatment’ network where, for instance, the institutional rules promote cooperative information or resource exchange versus a comparison network where institutional rules promote competition. The connections formed between members are tracked as the experiment unfolds, and the data are analyzed to evaluate the network dynamics in cooperative versus competitive institutional arrangements. 

\section{Discussion \& Conclusion}

While network research in public administration has grown tremendously in the past three decades, the vast majority of this research continues to be observational, lacking the experimental elements of intervention and control. There is a need move beyond descriptive work in order to address the what-works question. What can public managers do with information about networks to improve their functioning and performance? \textcite{Siciliano_and_Whetsell_2023} suggested network interventions as a framework for thinking about these issues, offering the classic example of \textcite{provan1995preliminary} who found that centralized networks tend to be more high performing. If a public manager took this insight seriously, what actions could they take to make the network more or less centralized? Finally, how would we know if such an intervention produced the expected outcome?

In the previous sections, we have presented questions, challenges, and models of network inference in relatively distinct containers. Yet these elements do overlap in unique ways. We argue that standard methodologies for handling threats to validity in causal inference will encounter new problems in the network context. Recently developed network methods may address some of these. Though we have certainly not exhaustively detailed all the problems, models, and their interactions, consider the following scenario. 

Take the case of a field experiment that seeks to determine the effects of an intra-organizational collaboration-based network intervention on processes and structures of network formation, where different departments are used as treatment versus comparison networks. After the design is implemented and the data are gathered, a stochastic actor oriented model could be used to estimate the the effect of the intervention while also accounting for endogenous network effects such as preferential attachment, reciprocity, and transitivity \parencite{steglich2012actor, steglich2022stochastic}. The implementation may be more or less difficult than this toy example suggest, but the general point is to leverage contemporary tools of network analysis to account for complex social social processes. Why do this? Accounting for these processes is not possible in the standard linear model, and ignoring them can seriously bias estimates. At the same time, the problem of network interference \parencite{leung2020treatment}, or spillovers, from treatment to comparison networks increases or decreases in severity depending on the degree to which organizational departments can be isolated from one another, e.g. physical, geographical, or virtual isolation. This presents unique problems for the way researchers construct control and comparison networks. It may even provide interesting opportunities to test for spillovers in a predictable manner, i.e. suppose we can anticipate spillovers in policy and program design in order to maximize implementation effectiveness. 

Inferential network models also enable the analysis of co-evolution of variables related to network formation and those related to outputs/outcomes. In principle, it is possible to estimate treatment effects on network formation as well as subsequent effects of formation on outcomes, all in a single model that also captures endogenous network processes. In summary, the use of inferential network methods on data collected with attention to experimental design elements could mark a significant step forward in combining insights from network methodologists and scholars of causal inference.

In conclusion, we have sought to address fundamental questions of causal inference within the context of social network data and analysis. While network analysis has flourished quite dramatically in public administration in recent decades, the vast majority of studies are observational in the sense that they lack the experimental elements of intervention, randomized control, or comparison networks. Given this lacuna, we discussed studies from a variety of disciplines that have made significant advances in network based causal inference. With this additional context in mind, we articulated a core set of problems specific to network inference and a typology of network inference models. Of course, there is good reason that these types of studies are far and few between, namely, that it is extremely difficult to gather temporal data across multiple networks. Hopefully, this article may serve as a guide for researchers seeking to establish causal inferences in the network setting. 

%\bibliographystyle{apacite}

%\bibliography{bib}
\printbibliography

\end{document}